
\texsis
\preprint
\singlespace
\def\greatvoid#1{ }

\def\ljets{$e^-e^+ \ra t \bar t \ra l^\pm + \hbox{jets}$}
\def\dilep{$e^-e^+ \ra t \bar t \ra b \bar{b} l^- l^+ \nu\bar{\nu}$}
\def\Ztt{\hbox{$Z\!-\!t-\!\bar t$}}
\def\gtt{\hbox{$\gamma\!-\!t-\!\bar t$}}
\def\Zee{\hbox{$Z\!-\!e^--\!e^+$}}
\def\gee{\hbox{$\gamma\!-\!e^--\!e^+$}}
\def\ee{$e^-e^+$}
\def\ra{\rightarrow}
\def\A{{\cal A}}

\def\half{{1 \over 2}}

\def\Lag{{\cal L}}

\def\del{\partial}

\def\journal#1&#2(#3)#4{{\unskip,~\sl #1\unskip~\bf\ignorespaces #2\unskip~\rm
(19#3) #4}}
\def\jour#1&#2(#3)#4{{\unskip ~\sl #1\unskip~\bf\ignorespaces #2\unskip~\rm
(19#3) #4}}
\def\uu{{\uparrow\uparrow}}
\def\ud{{\uparrow\downarrow}}
\def\du{{\downarrow\uparrow}}
\def\dd{{\downarrow\downarrow}}
\def\eett{{$e^-e^+\to t\bar{t}$}}

\pubcode{MSUTH 92/07}
\line{\hfil Revised}
\titlepage
\title{ A Probe of New Physics in Top Quark
Pair Production at $e^-e^+$ Colliders }
\author
G.~A. Ladinsky \ \ and \ \ C.--P. Yuan
{\it Michigan State University}
{\it Department of Physics and Astronomy}
{\it East Lansing, MI 48824-1116}
\endauthor

\abstract
{\baselineskip=1.0in
We describe how to probe new physics through examination of the form
factors describing the~\Ztt\ couplings via the scattering
process~\eett.
We focus on experimental methods on how the top
quark momentum can be determined and show how this
can be applied to select polarized samples
of $t\bar{t}$ pairs
through the angular correlations in the final state leptons.
We also study the dependence on the energy and luminosity of
an \ee\ collider to probe a CP violating asymmetry at the $10^{-2}$ level.}
\endabstract
\endtitlepage
%
%
\referencelist
\reference{baryon}
N.~Turok and J.~Zadrozny {\journal Phys.~Rev.&65 (90) 2331};
\hfill\break
M.~Dine, P.~Huet, R. Singleton and L.~Susskind
{\journal Phys.~Lett.&257B (91) 351};
\hfill\break
L.~McLerran, M. Shaposhnikov, N.~Turok and M.~Voloshin
{\journal Phys.~Lett.&256B (91) 451};
\hfill\break
A.~Cohen, D.~Kaplan and A.~Nelson{\journal Phys.~Lett.&263B (91) 86}
\endreference
\reference{ikm}
C.J.C.~Im, G.L.~Kane and P.J.~Malde, University of Michigan at Ann Arbor
preprint UM-TH-92-27
\endreference
\reference{peskin}
C.R.~Schmidt and M.E.~Peskin \journal Phys.~Rev.~Lett.&69 (92) 410
\endreference
\reference{weinberg}
S.~Weinberg \journal Phys.~Rev.~Lett.&63 (89) 2333;
\jour Phys.~Rev.&D42 (90) 860
\endreference
\reference{soni}
D.~Atwood and A.~Soni in \jour Phys.~Rev.~&D45 (92) 2405
\endreference
\reference{pisin}
see, e.g.,
P.~Chen \journal Phys.~Rev.&D46 (92) 1186
\endreference
\reference{*pisina}
P.~Chen, SLAC preprint SLAC-PUB-5914,
presented at the 9th International Workshop on Photon-Photon Collisions
(PHOTON-PHOTON '92), San Diego, CA, 22-26 Mar 1992
\endreference
\reference{*settles}
Ronald Settles in
Les Rencontres de Physique de la Vallee D'Aoste: Results and Perspectives in
Particle Physics, La Thuile, Italy, Mar 8-14, 1992 and at 27th Rencontre de
Moriond: Electroweak Interactions and Unified Theories, Les Arcs, France, Mar
15-22, 1992, and at 4th San Miniato Topical Seminar on the Standard Model and
Just Beyond, San Miniato, Italy, Jun 1-5, 1992
\endreference
\reference{peccei}
R.D.~Pecci and X.~Zhang{\journal Nucl.Phys.&B337 (90) 269};
\hfill\break
R.~Peccei, S.~Peris and X.~Zhang \journal Nucl.Phys.&B349 (91) 305
\endreference
\reference{tpol}G.L.~Kane, G.A.~Ladinsky and C.-P.~Yuan
\journal Phys.~Rev. &D45 (92) 124
\endreference
\reference{smcp}
C.~Jarlskog \journal Phys.~Rev.&D35 (87) 1685
\endreference
\greatvoid{
\reference{georgi}
M.~Chanowitz, H.~Georgi and
M.~Golden {\journal Phys.~Rev.~Lett.&57 (86) 2344};
\hfill\break
S.~Weinberg {\journal Phys.~Rev.&166 (68) 1568};
\hfill\break
S.~Coleman, J.~Wess and B.~Zumino {\journal Phys.~Rev.&177 (69) 2239};
\hfill\break
C.~Callan, S.~Coleman, J.~Wess and
B.~Zumino{\journal Phys.~Rev.&177 (69) 2247};
\hfill\break
S.~Weinberg {\journal Physica&(6A (79) 327};
\hfill\break
J.~Gasser and H.~Leutwyler {\journal Ann.~Phys.&158 (84) 142};
{\jour Nucl.Phys.&B250 (85) 465}
\endreference
\reference{counting}
A.~Manohar and H.~Georgi \journal Nucl.Phys.&B232 (84) 189
\endreference
}
\greatvoid{
\reference{fnal}
G.A.~Ladinsky \journal Phys.~Rev.&D46 (92) 3789
\endreference
\reference{trib}
G.L.~Kane, J.~Vidal and C.-P.~Yuan \journal Phys.~Rev.&D39 (89) 2617
\endreference
}
\reference{jcc}J.C.~Collins and G.A.~Ladinsky in
{\it Proceedings of the Polarized Collider Workshop}, The Pennsylvania
State University, 1990, ed. J.~Collins, S.~Heppelmann and R.~Robinett
(AIP Conference Proceedings No. 223, American Institute of Physics,
New York, 1991)
\endreference
\reference{minuit}
MINUIT, Application Software Group, Computing and Networks
Division, CERN, Geneva, Switzerland
\endreference
\reference{pawaii}
Talk presented by M.~Peskin at the Workshop on
Physics and Experiments with Linear $e^+e^-$ Colliders, Waikoloa,
Hawaii, 26-30 April 1993
\endreference
\reference{fadin}
E.A.~Kuraev and V.S.~Fadin, \jour Yad.~Fiz.&41 (85) 733
(\jour Sov.~J.~Nucl.~Phys.&41 (85) 466)
\endreference
\reference{chang}
D.~Chang, W.-Y. Keung and I.~Phillips, CERN preprint CERN-TH.6658/92
\endreference
\reference{skane}
G.L.~Kane, published in the {\it Proceedings of the 12th SLAC
Summer Institute on Particle Physics}, 23 July--3 August 1984,
Stanford, California
\endreference
\reference{dalitz}
R.H.~Dalitz and G.R.~Goldstein, {\jour Phys.~Lett.&B287 (92) 225;}
preprint OUTP-93-16P
\endreference
\reference{strikman}
L.L.~Frankfurt, et al. \journal Phys.~Lett.&B230 (89) 141
\endreference
\greatvoid{
}
\endreferencelist
%
%
\section{ Introduction }

Large CP violating effects are required to have the cosmological
baryon asymmetry produced at the weak phase transition.\cite{baryon}
If such effects exist, can they be probed in hadron and electron
colliders?\cite{ikm}
Schmidt and Peskin\cite{peskin} studied a CP violation
mechanism suggested by Weinberg\cite{weinberg} and concluded that
it is possible to probe a CP violating asymmetry in high--energy
hadron collider experiments via the production of top quark--antiquark pairs.
A detailed study on determining what observables are optimum for
measuring the form factors governing the
electric and magnetic dipole moments has been performed by
Atwood and Soni.\cite{soni}
In this paper it is shown that to probe the same level of CP violating
asymmetry at the Next Linear Collider (NLC) with $\sqrt{s}=500$ GeV
requires $3\times 10^{4}\, {\rm fb}^{-1}$ to produce about $2\times 10^7$
$t\bar t$ pairs, which is about a hundred times
larger than the proposed integrated luminosity.
(We use NLC to represent a generic $e^-e^+$ supercollider.\cite{pisin})
To slightly improve the measurement on the CP violating asymmetry
in~\ee\ collisions, we propose a different experimental observable
from the one suggested in \Ref{peskin} by measuring the polarization of the
$t$ and $\bar t$ in the process~\dilep. We show that the
efficiency of untangling the different polarizations for the $t$ and $\bar t$
increases as the energy of the~\ee\ collider increases at the cost of
decreasing the total event rate.  The net effect is that a $500\,$GeV
machine might do a better CP violation measurement than a $1\,$TeV
machine.

In present experimental data, the \Ztt\ coupling was measured via its
contribution to the decay of $Z \ra b \bar b$ and the mass shift of the $Z$
boson at the loop level.
It was argued that based on the electroweak chiral lagrangian analysis,
the form factors describing the~\Ztt\ interaction
can be different from the ones predicted by the
Standard Model (SM).\cite{peccei}
We propose measuring these form factors by
considering the polarization of the $t$ and $\bar t$.
We show how to identify the
polarization of the $t$ and $\bar t$ and how to
reconstruct the momenta of the $t$ and $\bar t$ via the decay mode
$ t \bar t \ra b l^+ \nu_l \bar b l^- {\bar \nu_l}$, where $l$
can be either an electron or a muon. We conclude that the form factors can be
tested up to a few percent for the NLC with a luminosity
of $50 \, {\rm fb}^{-1}$.
For simplicity, in this paper we follow the argument in \Ref{peccei} and
assume that the dominant non--standard form factors are only present
in the~\Ztt\ coupling and treat the couplings of the photon and $W^\pm$ boson
with fermions as described in the SM.

This paper is organized as follows.  In Section~2 we present the
interaction lagrangian and its corresponding helicity amplitudes which
describe the production of $t\bar{t}$ pairs via \ee\ annihilation
using general form factors.  There, we also present the Born level cross
sections in the Standard Model.
In Section~3 we review the requirements for discussing CP violation and
in Section~4 we examine how well the form factors that govern
the vector and axial vector interactions can be mneasured.
In Section~5 we discuss how to reconstruct the top quark momentum in
the $e^-e^+ \ra t \bar t \ra b \bar b l^- l^+ \nu\bar{\nu}$ and
$e^-e^+ \ra t \bar t \ra l^\pm +\hbox{jets}$ mode including
the effects of initial state radiation.  In Section~6
we discuss top quark polarization and how to select polarized event samples,
then in Section~7 we give our conclusions.

\section{The General Form Factors and Helicity Amplitudes}

\subsection{Conventions and Amplitudes}

In a recent paper\cite{tpol} we studied the most general
form factors for the coupling
of $t$ and $\bar t$ with either of the vector bosons $\gamma$ or $Z$.
Decomposing this interaction over the basis given by the Clifford
algebra in terms of the form factors $F^{R,L}_{1,2,3}$ yields
the general lagrangian,
$$\eqalign{
\Lag_{int}=& g\bigg\lbrack
Z_\mu\bar{t}\gamma^\mu(F_1^{Z(L)} P_-+F_1^{Z(R)} P_+)t
-{1\over v}\del_\nu Z_\mu\bar{t}\sigma^{\mu\nu}
(F_2^{Z(L)}P_-+F_2^{Z(R)}P_+)t	\cr
+&\del^\mu Z_\mu \bar{t}(F_3^{Z(L)}P_-+F_3^{Z(R)}P_+)t
+A_\mu\bar{t}\gamma^\mu(F_1^{\gamma (L)} P_-+F_1^{\gamma (R)} P_+)t
\cr
-&{1\over v} \del_\nu A_\mu\bar{t}\sigma^{\mu\nu}
(F_2^{\gamma (L)}P_-+F_2^{\gamma (R)}P_+)t
+\del^\mu A_\mu \bar{t}(F_3^{\gamma (L)}P_-+F_3^{\gamma (R)}P_+)t\bigg\rbrack
\cr}
\EQN intlag
$$
where $P_\pm ={1\over 2}(1\pm \gamma_5)$,
$i\sigma^{\mu\nu}=-{1\over 2}[\gamma^\mu,\gamma^\nu]$,
and $v=(\sqrt{2}G_F)^{-1/2}\sim 246\,$GeV.

Applying the Gordon decomposition, the vertex factors from the \Ztt\ and
\gtt\ interactions become
$$\eqalign{
\Gamma^\mu_Z =&{g\over 2}\bar{t}\lbrack\gamma^\mu(A_Z-B_Z\gamma^5)
+{p(t)^\mu -p(\bar{t})^\mu\over 2}(C_Z-D_Z\gamma^5)
+{p(t)^\mu +p(\bar{t})^\mu\over 2}(E_Z -F_Z \gamma^5)\rbrack t, \cr
\Gamma^\mu_\gamma=&
{g\over 2}\bar{t}\lbrack\gamma^\mu(A_\gamma -B_\gamma \gamma^5)
+{p(t)^\mu -p(\bar{t})^\mu\over 2}(C_\gamma -D_\gamma \gamma^5)
+{p(t)^\mu +p(\bar{t})^\mu\over 2}(E_\gamma -F_\gamma \gamma^5)\rbrack t,
\cr}
\EQN eone
$$
where
$$\eqalign{
A_{[\gamma,Z]}= & F_1^{[\gamma,Z](L)} +F_1^{[\gamma,Z](R)}
- {2m_t\over v}(F_2^{[\gamma,Z](L)}+F_2^{[\gamma,Z](R)}),  \cr
B_{[\gamma,Z]}=& F_1^{[\gamma,Z](L)}-F_1^{[\gamma,Z](R)}, \cr
C_{[\gamma,Z]}= & {2\over v}(F_2^{[\gamma,Z](L)}+F_2^{[\gamma,Z](R)}), \cr
D_{[\gamma,Z]}= & {2\over v}(F_2^{[\gamma,Z](L)}-F_2^{[\gamma,Z](R)}), \cr
E_{[\gamma,Z]}= & -2(F_3^{[\gamma,Z](L)}+F_3^{[\gamma,Z](R)}), \cr
F_{[\gamma,Z]}= & 2(F_3^{[\gamma,Z](L)}-F_3^{[\gamma,Z](R)}).
\cr}
\EQN formrel
$$
Since we ignore the masses of the incoming electron and positron, the
$F_3$ form factors in \Eq{intlag}
(i.e., the $E_{[\gamma,Z]}$ and $F_{[\gamma,Z]}$ in \Eq{formrel}, whose
vector coefficient is the four--momentum of either the
$\gamma$ or $Z$ vector boson in our problem)
do not contribute,
however, we note that current conservation for the photon demands that
$$
B_\gamma={-(p(t)+p(\bar{t}))^2\over 2m_t}(F_3^{\gamma (L)}-F_3^{\gamma (R)}).
\EQN currcons
$$

In the SM, at the Born level, the form factors are
$$\eqalign{
A_\gamma=&{4\over 3}\sin\theta_W,
\qquad\qquad\ \ \ \ \,
A_Z={1\over 2\cos\theta_W}(1-{8\over 3}\sin^2\theta_W),	\cr
B_\gamma=&0,
\qquad\qquad\qquad\qquad\ \
B_Z={1\over 2\cos\theta_W},	\cr
C_\gamma=&D_\gamma =0,
\qquad\qquad\qquad\phantom{.}
C_Z=D_Z=0 ,
\cr}
\EQN ethree
$$
where $\theta_W$ is the weak mixing angle.
Beyond the tree level, all of them except $D$, which controls the CP
violation, have contributions due to loop
corrections in the SM provided we ignore the small CP violating effects
which reside in the Yukawa couplings that govern the interactions between the
Higgs boson and the quarks. As argued in \Ref{smcp},
the CP violating amplitudes in the Cabbibo--Kobayashi--Maskawa model
are typically suppressed by a factor of order $10^{-12}$.

The \Zee\ and \gee\ vertices are given by
$$
\Gamma^\mu_{Zee}=ig\gamma^\mu(e^{Z}_LP_-+e^{Z}_RP_+)
\qquad\hbox{and}\qquad
\Gamma^\mu_{\gamma ee}=ig\gamma^\mu(e^{\gamma}_LP_-+e^{\gamma}_RP_+),
\EQN gzee
$$
where the SM values for these couplings are
$$e^Z_L ={1\over\cos\theta_W}(-\half +\sin^2\theta_W), \
e^Z_R ={1\over\cos\theta_W}\sin^2\theta_W,\ \hbox{and} \
e^\gamma_L=e^\gamma_R=-\sin\theta_W.
\EQN sixtwo
$$

The helicity amplitudes for \eett\ are represented by
$(h_{e^-},h_{e^+},h_{t},h_{\bar{t}})$,
where $h_{e^-}=-,+$ respectively indicates a left--handed
and a right--handed electron.
Apart from the common factor
$$
{2g^2E \over s-M_Z^2},
$$
with $s=4E^2$, the nonvanishing helicity
amplitudes from the diagram mediated by the $Z$--boson in \eett\ are,
in terms of these form factors $A_Z$, $B_Z$, $C_Z$ and $D_Z$,\cite{tpol}
$$\eqalign{
(-+--)_Z=&e_L\sin\theta_t[m_tA_Z-K^2 C_Z+E K D_Z],            \cr
(-+-+)_Z=&-e_L(1+\cos\theta_t)[EA_Z+K B_Z],            \cr
(-++-)_Z=&e_L(1-\cos\theta_t)[EA_Z-K B_Z],            \cr
(-+++)_Z=&e_L\sin\theta_t[-m_tA_Z+K^2 C_Z+E K D_Z],            \cr
(+---)_Z=&e_R\sin\theta_t[m_tA_Z-K^2 C_Z+EK D_Z],            \cr
(+--+)_Z=&e_R(1-\cos\theta_t)[EA_Z+K B_Z],            \cr
(+-+-)_Z=&-e_R(1+\cos\theta_t)[EA_Z-K B_Z],            \cr
(+-++)_Z=&e_R\sin\theta_t[-m_tA_Z+K^2 C_Z+E K D_Z] .           \cr
\cr}
\EQN etwo
$$
In the above formulas $E=\sqrt{s} /2 $ is half the center of mass energy
of the $e^-e^+$ annihilation and $K=\sqrt{E^2-m_t^2}$.
The nonvanishing Born helicity amplitudes from the photon propagator diagram,
$(h_{e-},h_{e+},h_{t},h_{\bar{t}})_\gamma$,
can be easily obtained from the results in \Eq{etwo} by setting
$e_L=e_R=-\sin\theta_W$ and replacing the $A,B,C,D$ form factors for
the $Z$ boson with those for the photon.  The common factor in this case is
$$
{2g^2E \over s}.
$$
The helicity amplitudes for the process \eett\ are therefore obtained by
summing the contributions from these two diagrams:
$$
(h_{e-},h_{e+},h_{t},h_{\bar{t}})=
2g^2E\bigg\lbrack{(h_{e-},h_{e+},h_{t},h_{\bar{t}})_Z\over{s-M_Z^2}} +
{(h_{e-},h_{e+},h_{t},h_{\bar{t}})_\gamma\over{s}}\bigg\rbrack .
\EQN sixfixa
$$
When calculating the cross section at the tree level,
a color factor of 3 should be included for the top quark pair production.
The spin average factor ${1 \over 2}{1\over 2}$ should also be included for an
unpolarized \ee.

\midtable{txsec}
\Caption
Born level production rates for \eett\
at $\sqrt{s}=500\,$GeV and $1\,$TeV using $m_t=140\,$GeV listed according to
$t$ and $\bar t$ polarizations.
\endCaption
\singlespaced
\ruledtable
\multispan3
\hfil CROSS SECTIONS FOR VARIOUS $t$ and $\bar t$ POLARIZATIONS \hfil\CR
$t\bar{t}$ Polarizations 	  \dbl$\sigma(\sqrt{s}=500\,$GeV) in fb
					   | $\sigma(\sqrt{s}=1\,$TeV) in fb\cr
$RR=LL$				  \dbl   37.3   |  2.69	\cr
$RL$				  \dbl   225    |  66.7	\cr
$LR$				  \dbl   342    |  104	\cr
$TT(\uu)_{in}=TT(\dd)_{in}$	  \dbl   93.6   |  29.8	\cr
$TT(\du)_{in}$			  \dbl   290    |  67.2	\cr
$TT(\ud)_{in}$			  \dbl   164    |  49.6	\cr
$TT(\ud)_\perp=TT(\du)_\perp$	  \dbl   131    |  32.5 \cr
$TT(\uu)_\perp=TT(\dd)_\perp$	  \dbl   190    |  55.7	\cr
$t$ is $R$, $\bar{t}$ unpolarized \dbl   262    |  69.4	\cr
$t$ is $L$, $\bar{t}$ unpolarized \dbl   379    |  107	\cr
top only is $T(\uparrow)_{in}$	  \dbl   258    |  79.4	\cr
top only is $T(\downarrow)_{in}$  \dbl   383    |  97.0	\cr
Unpolarized $t\bar{t}$		  \dbl   641    |  176
\endruledtable
\endtable

\subsection{Cross Sections in the Standard Model}

The production rates of polarized $t \bar t$ pairs at
500 GeV and 1 TeV~\ee\ colliders are tabulated in \Tbl{txsec}
for a $140\,$GeV top quark.
$LL$ (RR) denotes the production rate of
$t_L \bar t_L$ ( $t_R \bar t_R$ ) events, where $L$ labels a left--handed
helicity and $R$ labels a right--handed helicity.  Likewise,
$RL$ ($LR$) denotes the production rate of
$t_R \bar t_L$ ( $t_L \bar t_R$ ) events.
Besides the helicity cross sections, we have also included the cross
sections for transverse polarizations of the $t$ and $\bar t$ (indicated
by $T$).  In the first column of \Tbl{txsec}, the subscript $in$ or $\perp$
respectively indicate whether the cross sections refer to transverse
polarizations of the quark pair in the \eett\ scatter plane or perpendicular
to it.  If the arrows are (anti)aligned, the $t$ and $\bar t$ spins are
(anti)aligned.  An upward(downward) pointing arrow in the case of polarizations
perpendicular to the scatter plane means the spin is directed in the
$+\hat{\bf y}$ ($-\hat{\bf y}$) direction where
$\hat{\bf y}\equiv \hat{\bf z}\times \vec{\bf p}(t)$. (Recall the incoming
electron beam is moving in the $+\hat{\bf z}$ direction.)
An upward(downward) pointing arrow in the case of polarizations
in the scatter plane means the spin is directed
perpendicular to the top momentum and toward the $-\hat{\bf z}$
($+\hat{\bf z}$ ) direction.

Since $LR$ and $RL$ only depend on the form factors
$A$ and $B$, it is possible to untangle
the contributions from $A$ and $B$ by studying the polarized states
$t_L \bar t_R$ and $t_R \bar t_L$.
One important issue is how well the NLC can bound these form factors
for a given integrated luminosity.

If one can separate the polarization states of $LL$ and
$RR$ from the others, then one can measure the asymmetry
$$
\A_1={LL-RR\over LL+RR}
\EQN eseven
$$
or
$$
\A_2={LL-RR\over LL+RR+LR+RL} .
\EQN eeight
$$
These asymmetries are important for studying CP violation.

\section{CP Violation}

\subsection{Form Factors and CP Violation}

If the form factor $D$ in \Eq{eone} is not zero, the theory is CP
violating.\cite{tpol}
Specifically, from \Eq{etwo} it is apparent that the difference
between the $RR$ and $LL$ cross sections is linearly dependent on $D$ through
$$
(RR-LL)\propto
Re\Bigg\lbrack\bigg\lparen {A_Z\over s-m_Z^2} +
		{A_\gamma\over s}\bigg\rparen D_Z^*\Bigg\rbrack,
\EQN cpviol
$$
where the ``$*$'' indicates complex conjugation and
in this equation the photon is assumed to preserve its
SM behavior, namely, $D_\gamma=0$.
{}From the helicity amplitudes listed in
\Eq{etwo}, one can construct a quantity sensitive to CP violation,
$\A_1={LL-RR\over LL+RR}$, to measure the real part of $D$.\cite{peskin}
The imaginary part of $D$ can be examined by studying the transverse
polarization of the top quark pairs perpendicular to the scatter
plane.\cite{tpol}
The advantage of being able to measure the polarization states
of the top quark pair is that one can
measure $\A_1$ instead of $\A_2={LL-RR\over LL+RR+LR+RL}$ which may lose
some of its effectiveness due to the dilution acquired by having a larger
denominator from the $RL$ and $LR$ events when studying
the CP violation effects.

\figure{feight}
\vskip 1.0in
\Caption
This plot displays the variation of the asymmetry $\A_1$ in
\Eq{eseven} with the mass of the Higgs boson for Weinberg's model using
$\sqrt{s}=500,1000\,$GeV and $m_t=140\,$GeV.
\endCaption
\endfigure

In Weinberg's model there is a dependence on the mass of the Higgs boson that
enters in the form factor $D$,
$$
\hbox{Re}\{ D_{[\gamma,Z]} \}= 2\hbox{Im}\{ Z_2 \}
A^{\hbox{Born}}_{[\gamma,Z]}
\bigg\lparen {m_t^4 \over 4\pi v^3s\beta }\bigg\rparen
\bigg\lbrack 1-{m_H^2\over s\beta^2}\ln (1+{s\beta^2\over m_H^2})\bigg\rbrack,
\EQN wein
$$
where $\beta=\sqrt{1-4m_t^2/s}$ and
$A^{\hbox{Born}}_{[\gamma,Z]}$ refers to the SM values at the Born level
for $A_\gamma$ and $A_Z$ given in \Eq{ethree}.  For models with two or
more Higgs doublets Weinberg showed that $2\hbox{Im}\{ Z_2 \}\le \sqrt{2}$
with a reasonable choice of Higgs vacuum expectation values.
When we compute $\A_1$ for a $500\,$GeV machine with $m_t=140\,$GeV
(see \Fig{feight}), we find $\A_1=-1.57\%$ for $m_H=100\,$GeV
and $\A_1=-0.43\%$ for $m_H=1\,$TeV.  The percentage decrease in the magnitude
of the CP violating asymmetry is less for a heavier top quark, but the
general trend is a decrease in $|\A_1|$ as the Higgs boson mass gets larger.

\figure{fnine}
\vskip 1.0in
\Caption
This plot displays the variation of the asymmetry $\A_1$ in
\Eq{eseven} with $\sqrt{s}$ for Weinberg's model using $m_t=140\,$GeV
and $m_H=0.1,0.4,1.0\,$TeV.
\endCaption
\endfigure

Using Weinberg's model, it is apparent that there is also a variation
of the CP violating asymmetry with the beam energy of the collider.
In \Fig{fnine} we plot $\A_1$ versus the center of mass energy for
Higgs boson masses of $100$, $400$ and $1000\,$GeV using
$m_t=140\,$GeV.  Comparing a $1\,$TeV machine with a $500\,$GeV
machine we see that the $\A_1$ asymmetry value is slightly larger
in magnitude for
the higher energy machine.  It is important to note, however, that the
$RR$ and $LL$ event rates, which are the ones used for studying CP
violation, are smaller by an order of magnitude for the $1\,$TeV
machine when compared with the $500\,$GeV machine (see \Tbl{txsec}).
Because of this reduction in statistics, the $500\,$GeV machine is
preferred for testing CP violation.

The question is,``What luminosity is needed to
measure the CP violating asymmetry $\A_1$ at the $10^{-2}$ level at
the NLC?''

\subsection{Requirements for studying CP Violation}

To test CP violation, we use the dilepton mode,
$$
e^-e^+ \ra t \bar t \ra b\bar{b}l^- l^+\nu\bar{\nu},
\EQN lmode
$$
which has a branching ratio of $2{1\over 9}{1\over 9}={2\over 81}$
for $l=e,\mu$.
At the Born Level in the Standard Model the unpolarized cross section for
\eett\ with $m_t=140\,$GeV is (see \Tbl{txsec})
$$
\sigma (e^-e^+\to t\bar{t})=
\cases{641\,\hbox{fb} &for $\sqrt{s}=500\,$GeV,\cr
       176\,\hbox{fb} &for $\sqrt{s}=1  \,$TeV,\cr}
\EQN xsecun
$$
while the cross section for the production of $t$ and $\bar t$ with
identical helicities is
$$
\sigma (e^-e^+\to t_L\bar{t}_L)=\sigma (e^-e^+\to t_R\bar{t}_R)=
\cases{37 \,\hbox{fb} &for $\sqrt{s}=500\,$GeV,\cr
       2.7\,\hbox{fb} &for $\sqrt{s}=1  \,$TeV.\cr}.
\EQN xsecll
$$

To measure an asymmetry $\A_1\sim 10^{-2}$ requires that the number of
events, $N$, be large enough so that the statistical error
in the measurement, ${1\over\sqrt{N}}$ is less than $10^{-2}$.  For the
optimistic value of ${1\over\sqrt{N}}\approx 10^{-2}$ we can estimate the
luminosity required to observe this CP violating asymmetry,
$$
{\cal L}={81N\over 2 \epsilon
[\sigma(e^-e^+\to t_R\bar{t}_R)+\sigma(e^-e^+\to t_L\bar{t}_L)]},
\EQN lumcp
$$
where $\epsilon$ is the efficiency of finding $LL$ or $RR$ events.  Assuming
we are able to solve for the kinematics of the dilepton mode in 70\%
of the events (see Section~5) and the acceptance of the kinematic cuts in
selecting $LL$ or $RR$ events is ${1\over 4}$ (see Section~6),
then $\epsilon\approx 0.18$.
This implies the required luminosity to observe CP violation
through $\A_1\sim 10^{-2}$ in top quarks of $m_t=140\,$GeV
is ${\cal L}\approx 3\times 10^4\,\hbox{fb}^{-1}$ at $\sqrt{s}=500\,$GeV
(i.e., about $2\times 10^7$ $t\bar t$ pairs are required).

\section{How Well Can Form Factors be Measured at the NLC?}

The bounds determined from the experimental data,
which limit the size of the form factors governing the top quark interaction
with the vector boson,
do not eliminate the possibility of nonuniversal couplings,
as discussed by Peccei and Zhang,\cite{peccei}
making it important to test the couplings as completely as we can.
When determining form factors for an s-channel process where the
intermediate state is a vector particle, one benefits by the
information conveyed by the variation in the angular distribution of
the final state.\cite{jcc} From \Eq{etwo} it is clear the $t\bar t$
helicity states are produced with a polar angle distribution of
either $\sin^2\theta$ or $(1\pm\cos\theta)^2$, so noting that the form
factors carry no angular dependences, it is possible to use the methods
described above to solve for the direction of motion for the top quark
and then use the differential cross section as described through the
parameters $c_0,c_+,c_-$ by
$$
{d\sigma\over d\cos\theta}={\sqrt{E^2-m_t^2}\over 32\pi sE}
[c_0\sin^2\theta+c_+(1+\cos\theta)^2+c_-(1-\cos\theta)^2 ].
\EQN decomp
$$
{}From \Eq{etwo} and \Eq{decomp} we obtain
$$\eqalign{
c_{0}=& c_{0+}+c_{0-}, \cr
c_{0\pm}=&	6(g^2E)^2
\Bigg\{ \bigg| e_L^Z{m_tA_Z-K^2C_Z\pm EKD_Z\over s-m_Z^2}
+e_L^\gamma {m_tA_\gamma-K^2C_\gamma \pm EKD_\gamma\over s}\bigg| ^2	\cr
&\qquad\qquad
+\bigg| e_R^Z{m_tA_Z-K^2C_Z\mp EKD_Z\over s-m_Z^2}
+e_R^\gamma {m_tA_\gamma-K^2C_\gamma\mp EKD_\gamma\over s}\bigg| ^2 \Bigg\}
,\cr
c_\pm=& 3(g^2E)^2
\Bigg\{
\bigg| e_L^Z{EA_Z\pm KB_Z\over s-m_Z^2}+e_L^\gamma{EA_\gamma\pm KB_\gamma
\over s}\bigg| ^2	\cr
&\qquad\qquad\qquad\qquad\qquad
+\bigg| e_R^Z{EA_Z\mp KB_Z\over s-m_Z^2}+e_R^\gamma{EA_\gamma\mp KB_\gamma
\over s}\bigg| ^2
\Bigg\}.
\cr}
\EQN cparam
$$

\figure{fone}
\vskip 1.0in
\Caption
In $d\sigma /d\cos\theta$ we see the dependence of the production
rates for polarized $t\bar{t}$ states on the top quark polar angle.  This angle
$\theta$ is measured with respect to the incoming $e^-$ beam which is taken to
move along the $+\hat{z}$ direction.
\endCaption
\endfigure

Untangling the form factors from such distributions is not trivial
for various reasons.
For instance, it is possible for there to be cancellations among the form
factors making deviations from the SM difficult to observe.
One result of these
complications is that given a certain polar angle distribution described by
a finite number of data points carrying some doubt in precision,
there are different sets of values for the form factors that can satisfy
the constraints such a distribution gives.  With this in mind, nevertheless,
we can ask how well experiments at the NLC might be able to constrain the
form factors.

\figure{ffive}
\vskip 1.0in
\vskip -1in
\Caption
These plots present the angular correlation between the charged
leptons produce from the polarized $t$ and $\bar t$ decays in \dilep.
Boosting to the individual center of mass frame for the decays of the $t$
and $\bar t$, we plot the cosine of the angle between the momentum of each
charge lepton and the boost axis as defined by the top momentum.  Note how each
combination of $t$,$\bar t$ helicities (RR,LL,RL,LR)
occupy separate quadrants.
\endCaption
\endfigure

\figure{fsix}
\vskip 1.0in
\Caption
The distribution for the production of polarized $t\bar{t}$ pairs,
$d\sigma /d[(1-r)/(1+r)]$ vs. $(1-r)/(1+r)$ for $r\equiv E(l^+)/E(l^-)$ when
angular cuts are applied to the final state $l^+$ and $l^-$ in the
respective decay frames of the $t$ and $\bar{t}$.
\endCaption
\endfigure

\midtable{tone}
\Caption
The upper and lower bounds indicate the statistical
range within which we can
determine the form factors from the angular distribution of the top quark
in \dilep\
given a $50\,\hbox{fb}^{-1}$ luminosity at the NLC within a $68\%$ confidence
level.
\endCaption
\singlespaced
\ruledtable
\multispan4\hfil FORM FACTOR BOUNDS FOR THE NLC \hfil\CR
Form Factor       \dbl   SM VALUE |  Upper   |  Lower   \cr
$F_1^{Z(L)}$      \dbl ~0.395     | ~0.419   | ~0.370   \cr
$F_1^{Z(R)}$      \dbl -0.175     | -0.153   | -0.197   \cr
$F_2^{Z(L)}$      \dbl ~0.0       | ~0.013   | -0.009   \cr
$F_2^{Z(R)}$      \dbl ~0.0       | ~0.013   | -0.009
\endruledtable
\endtable

\midtable{ttwo}
\Caption
The upper and lower bounds indicate the statistical
range within which we can
determine the form factors from the angular distribution of the top quark
in \dilep\
given a $50\,\hbox{fb}^{-1}$ luminosity at the NLC within a $90\%$ confidence
level.
\endCaption
\singlespaced
\ruledtable
\multispan4\hfil FORM FACTOR BOUNDS FOR THE NLC \hfil\CR
Form Factor         \dbl SM VALUE  | Upper     |  Lower   \cr
$F_1^{Z(L)}$        \dbl ~0.395    | ~0.473    | ~0.311   \cr
$F_1^{Z(R)}$        \dbl -0.175    | -0.107    | -0.254   \cr
$F_2^{Z(L)}$        \dbl ~0.0      | ~0.084    | -0.024   \cr
$F_2^{Z(R)}$        \dbl ~0.0      | ~0.084    | -0.024
\endruledtable
\endtable

We now proceed to ask how well the fit to the angular distribution can
do to get the form factors.
We show in \Fig{fone} the lowest order
SM production rates for polarized $t \bar t$ pairs as a function of the
scattered polar angle, $\theta_t$, measured in the center--of--mass frame of
the $t \bar t$ pair at the NLC.
In \Tbl{tone} and \Tbl{ttwo}
we present bounds that represent a $68\%$ and a $90\%$ confidence level
on the range for determining the form factors for the \Ztt\ interaction.
These bounds were obtained using MINUIT\cite{minuit}
to fit the polar angle distribution
of the top quark as generated for the NLC ($\sqrt{s}$=500 GeV, with
$50 \, {\rm fb}^{-1}$ integrated luminosity) with no constraints on the
kinematics.
We expect 30,000 $t\bar{t}$ events,
but if we focus on the mode where the two $W$'s decay leptonically to
$\mu^\pm$ or $e^\pm$, there is a branching ratio reduction to 1,500 events.
These 1,500 events were collected into twenty bins for the fit.
Allowing only one form factor to vary at a time, the results presented in
\Tbl{tone} (\Tbl{ttwo}) indicate that within the $68\%$ ($90\%$)
confidence limit, it should be
possible to find $F_1^L$ to within about $6\%$ ($20\%$), while
$F_1^R$ can only be known to within roughly $11\%$ ($40\%$).
The $F_2$ form factors are zero in the SM, and the fit indicates that their
values can be known to within about $0.0\pm 0.01$ ($0.0^{+0.08}_{-0.02}$)
at a confidence level of $68\%$ ($90\%$).

By taking our knowledge of
the polarization behavior to untangle some of the form factors before
comparing them with data, it may be possible to improve these bounds.
In particular from \Eq{etwo} it is apparent that the $t\bar t$
states of opposite helicity only depend on $A$ and $B$.  By making cuts
(to be discussed in Section~6),
we can isolate $RL$ and $LR$
contributions and investigate $A$ and $B$ from this data subset
with a minimum contamination from the $LL$ and $RR$ states.  Afterwards,
we can proceed to the $RR$ and $LL$ contributions to study $C$ and $D$.
Performing the same procedures as we did with the unpolarized cross section,
we take 750 events from the unconstrained $LR$ sample and use MINUIT to fit for
$F_1^L(Z)$ and find $0.395\pm 0.014$ at a confidence level of $68\%$,
which is roughly two percent better
than we did with the unpolarized distribution.
Further improvement in the form factor determination can come from the
increased statistics (by about a factor of 7)
obtained by including the events from
\hbox{$e^-e^+\to t\bar{t}\to l^\pm +\hbox{jets}$} in the analysis.
For the $l^\pm +\hbox{jets}$ mode, where the branching ratios
take $9,000$ unpolarized events from the original $30,000$,
the results indicate that within the $68\%$ ($90\%$)
confidence limit, it should be
possible to find $F_1^{(L)}$ to within about $3\%$ ($8\%$),
while
$F_1^{(R)}$ can be known to within roughly $5\%$ ($18\%$).
In this case the $F_2$ form factors
can be known to within about
$0.0\pm 0.004$ ($0.0\pm 0.02$) at a confidence level of $68\%$ ($90\%$).

More information can be gained by focusing on polarization
states for the top quarks besides pure helicity states.
For example, the interference terms that
result between the product of the imaginary parts of the amplitudes
with the real parts do not exist in the total rate expressed by
\Eq{decomp} and \Eq{cparam}, but
they do appear in the rate for top quark spins perpendicular to the scatter
plane.  To see this, note that the polarization asymmetry comparing
top spins pointing in opposite directions perpendicular to the scatter
plane is proportional to
$$
\sum_{h_{e-},h_{e+},h_{\bar{t}}}
\hbox{Im}[(h_{e-},h_{e+},h_{t}=+,h_{\bar{t}})
(h_{e-},h_{e+},h_{t}=-,h_{\bar{t}})^\dagger].
\EQN xperp
$$
Thus, angular correlations and decay plane correlations will further
improve the determination of the form factors $F_1$ and $F_2$.

\figure{ften}
\vskip 1.0in
\Caption
Same as \Fig{fone} except for a purely left--handed $e^-$ beam.
\endCaption
\endfigure

Even more useful, however, may be the purity of the $t\bar{t}$ polarization
which is obtained by polarizing the electron beam.\cite{pawaii}
The majority of the cross
section is comprised of left--handed top quarks from the start.  By selecting
a left--handed polarization for the electron beam, helicity conservation
suppresses the production of right--handed top quarks that move along the
initial electron direction of motion.  This can be seen in \Fig{ften}
where for each top quark helicity combination we plot
$d\sigma/d\cos\theta$ vs. $\cos\theta$.  What this provides is a cleaner
measure of the form factor dependence $EA+KB$, as apparent from the
amplitudes shown in \Eq{etwo}.  Similarly, a right--handed
polarization for the electron beam
will provide a cleaner sample of right--handed top quarks
where the form factor probed is $EA-KB$.

\section{How to reconstruct the Kinematics of the Top Quarks}

In reality, the initial state electron or positron at high energies
will radiate photons along the beam axis either due to initial state
radiation (ISR) during the $e^+e^-$ interaction or due to the interaction
with the classical electromagnetic field formed by having a dense bunch
of charged particles in the interaction region (beamstrahlung).
Such radiation tends to move along the beam direction
such that neither the center--of--mass energy nor the boost
of the $t \bar t$ pair is known, complicating the analysis
discussed in the previous section.
Despite this, the effects of initial state radiation are such that
most of the time the center--of--mass energy of the $t \bar t$ pair is
close to the beam energy.
Furthermore, it is possible to design experiments such that the
beamstrahlung is minimized, so we only consider the bremsstrahlung
effect.  Nevertheless, it is still possible to determine
the momentum of the top quark.

We direct our attention to solving for the top
quark momentum for two cases:  The first case will be the mode where
the top quark decays leptonically while the top antiquark decays
hadronically, $e^-e^+\to t\bar{t}\to bl^+\nu\bar{b}qq'$;
and the second case will be the mode where both top quarks decay leptonically,
$e^-e^+\to t\bar{t}\to bl^+\nu\bar{b}l^-\bar{\nu}$.
For the initial state radiation we use the distribution for getting
an $e^-$ (or $e^+$) with a fraction $z$ of its original beam energy, given by
Kuraev and Fadin\cite{fadin} to order $\lambda$,
$$
D(z,s)={\lambda\over 2}(1-z)^{{\lambda\over 2}-1}(1+{3\over 8\lambda})
	-{1\over 4}\lambda(1+z),
$$
where $\lambda\equiv{2\alpha\over\pi}(\ln {s\over m_e^2}-1)$ for original
center of mass energies of $\sqrt{s}$ and electron mass $m_e$.
In Section~6.2 we show that there was little difficulty in
associating the proper bottom flavored quark with the proper top flavored
quark, so we will drop the concern about this ambiguity for this part of
the discussion.

\subsection{$e^-e^+\to t\bar{t}\to bl^+\nu\bar{b}qq'$}

When one top quark decays hadronically and the other decays leptonically,
the only information we are missing to describe the momenta and energies
of the event completely is the component of the neutrino momentum that
moves along the beam axis.  The transverse momentum of the neutrino
is known by applying conservation of momentum.
Next, we apply a familiar formula which provides
a doubly degenerate solution for the longitudinal momentum of the
neutrino,
$$
p_z(\nu )={a\pm\sqrt{a^2-(1-b^2)(a^2-b^2 p_T^2(\nu ))}\over 1-b^2}.
\EQN pznu
$$
In the above formula,
$$
a={-(\vec{\bf p}_T(bl^+)\cdot\vec{\bf p}_T(\nu)+\half(m_t^2-M_{bl^+}^2))\over
p_z(bl^+)},     \qquad
b={E(bl^+)\over p_z(bl^+)},
\EQN pznuvars
$$
where $\vec{\bf p}(\nu )=(\vec{\bf p}_T^{}(\nu ),p_z(\nu ))$ is the
momentum of the neutrino separated into directions respectively
transverse and along
the $Z$--axis, while $p(bl^+)=(E(bl^+),\vec{\bf p}(bl^+))$
describes the four--momentum of the two--particle system of mass
$p(bl^+)\cdot p(bl^+)=M_{bl^+}^2$ composed of the
bottom quark and the charged lepton.
With the idea that polarization studies are a means of investigating
top quark properties {\it after} its discovery, the top quark mass is
assumed to be known and its decay width narrow.
In our Monte Carlo study, the widths of the $W$ bosons have been included
for event simulation, so this computation
was performed by scanning a range of values for $m(W^-)$ and $m(W^+)$
about the peak value of $m_W$
and accepting the first solution that was found.
Attempting to account for some detector
inefficiency, this procedure also included
a gaussian smearing of the momentum for the $b$ and $\bar b$ with
$\Delta E(b)/E(b)=0.5/\sqrt{E(b)}$ and for the visible leptons with
$\Delta{E(l^\pm)}/{E(l^\pm)}=0.15/\sqrt{E(l^\pm)}$.
(More detailed discussion is given in Section~6.2.)
Random chance yields a $50\%$ probability for selecting the
correct solution for $p_z(\nu)$, but we need to do better than that.
What needs to be resolved then, is how to remove this degeneracy in the
$z$--momentum of the neutrino.

\figure{isrshat}
\vskip 1.0in
\Caption
The center--of--mass energy of the $e^-e^+$ annihilation when
ISR is taken into account.
\endCaption
\endfigure

\midtable{tthree}
\Caption
The efficiency in solving for the top quark momentum
using initial state radiation, smearing, and a Breit--Wigner width
on the $W$--bosons is described here by three values:
the percentage of solveable configurations; the
cosine of the angle between the true top quark direction and the direction
given by solving the kinematics ($\zeta\equiv\cos(t_{true},t_{solved})$);
the root mean squared of $\zeta$.
\endCaption
\singlespaced
\ruledtable
\multispan4
\hfil ~SOLUTION EFFICIENCY(unknown $m_W$, smearing and ISR included)~ \hfil\CR
$l^++jets$                  \dbl \%Solved |$<\zeta>$|$<\zeta^2>-<\zeta>^2$ \cr
Best $m_W$                  \dbl 84       | 0.978   | 0.090 	           \cr
Best $E_{cm}$               \dbl 72       | 0.981   | 0.073 	           \CR
$l^+l^-b\bar{b}\nu\bar{\nu}$\dbl\%Solved  |$<\zeta>$|$<\zeta^2>-<\zeta>^2$ \CR
Best $E_{cm}$               \dbl 68       | 0.957   | 0.142
\endruledtable
\endtable

We have two straightforward means at our disposal for selecting
the best neutrino solution.  In one case we take advantage of the
fact that the width of the $W$--boson is narrow with respect to its
mass and select the $p_z(\nu)$ that gives a value for $(p(\nu)+p(e^+))^2$
that is closest to $m_W^2$.  In the second case we take advantage of the
feature that the initial state radiation is strongly peaked for soft
emissions, as shown in \Fig{isrshat}, and select the $p_z(\nu)$ that gives
a value for the center--of--mass
energy that is closest to the original beam energy.  To measure how these
two methods compare for determining the top quark momentum, we tabulate three
values in \Tbl{tthree}.  First, as a matter of statistics, we give the
percentage of the time this method was able to find a solution.  Selecting the
best $m_W$ provided the largest number of solutions where only $16\%$
of the cross section was lost while choosing the smallest loss of energy
through radiation failed to find a solution $28\%$ of the time.  What
may be more important in polarization studies
than losing an extra $12\%$ of the events, however,
is getting a more precise measure for the direction of the top quark.
Averaging the cosine of the angular separation between the true top quark
direction and the direction obtained by selecting the solution to the quadratic
formula for $p_z(\nu)$, which we label
$\zeta\equiv\cos(t_{true},t_{solved})$,
it was found that the minimum radiation method
had a value for $\zeta$ slightly closer to unity.
Not only is the average of this angle important, but also the spread
of this distribution, $<\zeta^2>-<\zeta>^2$, where it was better
to accept those solutions which provided the value of $M(t\bar{t})$ closest to
the original beam energy.

With either of these two methods, there is little loss in statistics
and very little change in the polar angle distribution of the top
quarks, making it possible to study the form factors through the polar
angle distribution of the top quark.

\subsection{$e^-e^+\to t\bar{t}\to bl^+\nu\bar{b}l^-\bar{\nu}$}

When the top quark and antiquark decay leptonically and both the boost
and the mass of the $t\bar{t}$ pair are unknown, there is a great deal
of information that must be extracted out of the measurement of the
four visible particles ($b,\bar{b},l^+,l^-$ for $l=e,\mu$).
Detailed discussion on such
kinematics can be found in \Ref{dalitz}.
The four--momenta of both the $\nu$ and
$\bar{\nu}$ must be determined using the center of mass energy as an
unknown.  This means we have eight unknowns.  Generalizing the
procedures that have been discussed here ealier, however, we find that
we can solve this system exactly, albeit within an eight--fold
degeneracy at most.

Determining the neutrinos' momenta is accomplished by utilizing the
fact that this process contains the decay of four particles
whose masses either are or will be known
and whose decay widths are narrow.  From the
start, we have the conservation of transverse momentum,
$$
\vec{\bf V}_T+\vec{\bf p}_T(\nu )+\vec{\bf p}_T(\bar{\nu})={\bf 0},
\EQN transsum
$$
where we define the four-vector for the sum of the
energies and momenta of the visible particles to be
$$
V\equiv (V^{(0)},\vec{\bf V})=p(b)+p(\bar{b})+p(e^+)+p(e^-).
\EQN vee
$$
The conservation of transverse momentum
provides two equations in pursuit of solving for the $p(\nu)$ and
$p(\bar{\nu})$.  Next, from the the decays of the $W$'s and $t$'s and the
masses of the $\nu$ and $\bar \nu$
we have six more relations, two of which are nonlinear:
$$\eqalign{
\hbox{Top quark mass:}&\ \ m(t)^2=(p({b})+p({\nu})+p(e^+))^2, \cr
\hbox{Top antiquark mass:}&\ \
                      m(\bar{t})^2=(p(\bar{b})+p(\bar{\nu})+p(e^-))^2, \cr
\hbox{$W^+$ mass:}&\ \ m(W^+)^2=(p(e^+)+p({\nu}))^2, \cr
\hbox{$W^-$ mass:}&\ \ m(W^-)^2=(p(e^-)+p(\bar{\nu}))^2, \cr
\hbox{neutrino mass:}&\ \ 0=E(\nu)^2-\vec{\bf p}(\nu)\cdot\vec{\bf p}(\nu), \cr
\hbox{antineutrino mass:}&\ \
0=E(\bar{\nu})^2-\vec{\bf p}(\bar{\nu})\cdot\vec{\bf p}(\bar{\nu}).
\cr}
\EQN system
$$
In this study we have used the narrow width approximation for
the top quark and top antiquark.

These
relations provide two coupled quadratic realtions in $E(\bar{\nu})$ and
$E({\nu})$.  Combining these two quadratic equations provides a quartic
equation in $E({\nu})$ which can be solved exactly, yielding a four--fold
degeneracy at most for the neutrino energy.  Substituting each solution
for the neutrino energy, one at a time, into either of the two quadratic
relations in $E(\bar{\nu})$ and $E({\nu})$ provides two solutions
for $E(\bar{\nu})$ for each $E({\nu})$.  The result is at most eight pairs of
solutions for ($E(\bar{\nu})$,$E({\nu})$).  Since the remaining equations
are linear, the solutions for the momentum components are unique for each
given ($E(\bar{\nu})$,$E({\nu})$) pair.  What remains is to decide which
solution is the correct one.

Since the initial
state radiation is peaked for soft emissions, as shown in \Fig{isrshat},
the method we use
for choosing among the degenerate solutions in this top decay mode is to
select the set of ($E(\bar{\nu})$,$E({\nu})$) that provides a value of
$M(t\bar{t})$ closest to the original beam energy.
In \Tbl{tthree}, we see that
we were able to obtain solutions for $68\%$ of the cross section
with $<\zeta>=0.957$ and $<\zeta^2>-<\zeta>^2=0.142$.

\figure{betadis}
\vskip 1.0in
\Caption
The boost, $d\sigma /d\beta$ vs.
$\beta={\vec{\bf p}(t)+\vec{\bf p}(\bar{t})\over E(t)+E(\bar{t})}$,
as generated by ISR.
\endCaption
\endfigure

\figure{shcurve}
\vskip 1.0in
\Caption
The angular distribution $d\sigma /d\cos\theta$ vs. $cos\theta$
as dictated by the top momentum provided by the Monte Carlo
compared against the top momentum found when solving the kinemtics
incuding the effects of ISR.
\endCaption
\endfigure

\subsection{Initial State Radiation}

The effects of ISR on the angular distribution of the top quarks is minimal.
To demonstrate this, we show in \Fig{isrshat} and \Fig{betadis} that
the initial state
radiation is soft, keeping the energy of the process quite close to the
beam energy for a majority of the events.
The $t\bar{t}$ system is hardly boosted as illustrated in \Fig{betadis},
where
$\beta={\vec{\bf p}(t)+\vec{\bf p}(\bar{t})\over E(t)+E(\bar{t})}$.
When we take events generated
by our Monte Carlo and compare the angular distributions between the
kinematics obtained by solving the eight simultaneous equations
and the actual value generated by the program, we find only a slight
difference, as shown in \Fig{shcurve}.

With the $t$ and $\bar t$ momenta determined, it becomes feasible to
select a sample of $t\bar{t}$ pairs that is strongly biased to a
particular polarization state.

\section{Determining the Polarizations of the $t$ and the $\bar t$}

\figure{ftwo}
\vskip 1.0in
\vskip -3in
\Caption
Helicity arguments control the preferred moving direction for the
$e^+,e^-$ as they are produced in the respective rest frames
of the $t,\bar{t}$ decays ($t\to be^+\nu$, $\bar{t}\to \bar{b}e^-\bar{\nu}$).
The subscript $R$ ($L$) indicates right--handed (left--handed) quark helicity.
\endCaption
\endfigure

\figure{fthree}
\vskip 1.0in
\Caption
$E(l^+)$ and $E(l^-)$ distributions for \eett\ when both final
state top quarks have identical helicities.
\endCaption
\endfigure

\subsection{Polarized $t\bar{t}$ Production}

Here we consider the dilepton mode.
We showed in \Ref{tpol} that the polarization of the $t$ (and the $\bar t$)
can be self--analyzed from the decay $t \ra b W^+\ra bl^+ \nu_{l}$
($\bar t \ra \bar b W^-\ra \bar{b}l^- \bar \nu_{l}$).
For a heavy top quark, the preferred moving direction of the $l^+$ and
the $l^-$ in
the center--of--mass frame of the $t \bar t$ pair are shown in \Fig{ftwo}.
Because of the correlations,
it is possible to distinguish different polarization states of the
$t \bar t$ pairs by the energy distribution, $d\sigma/dE(l)$,
of the leptons $l^+$ and $l^-$.\cite{peskin}\cite{chang}\cite{skane}
Because the asymmetry in $E(l^+)$ and $E(l^-)$ is magnified when the $t$ and
the $\bar t$ are boosted, we anticipate that it becomes easier to distinguish
different polarization states of the $t \bar t$ pair when the energy of
the~\ee\ collider increases,  but it must be kept in mind that
the production rate of $t \bar t$ decreases with higher beam energies.
At the NLC, the energy distributions of the $l^+$ and $l^-$ are shown
in \Fig{fthree}.

\figure{ffour}
\vskip 1.0in
\Caption
The distribution for the production of polarized $t\bar{t}$ pairs,
$d\sigma /d[(1-r)/(1+r)]$ vs. $(1-r)/(1+r)$ for $r\equiv E(l^+)/E(l^-)$.
\endCaption
\endfigure

As discussed in the previous section, $\A_1$ is cleaner than $\A_2$ because
it does not contain the relatively large contributions from $LR$ and $RL$
in its denominators. Obviously, this can be useful only if one can separate
the various polarization states for the $t \bar t$ pairs.
One method to do this which applies the physics contained in the energy
distributions of \Fig{fthree} is to make a selection of events using the
energy ratio $r=E(l^+)/E(l^-)$.
The events with $r \gg 1$ are mainly $RR$ events, $r \ll 1$ are mainly $LL$,
and $r\sim 1$ characterizes both $RL$ and $LR$ events.
In order not to induce the CP asymmetry while applying the kinematic cut on
this variable $r$, it is necessary to make a {\it symmetric} cut on $r$. For
instance, one can select the $RR$ events by requiring $r > 3$ and the $LL$
events by requiring $r < 1/3$.
In \Fig{ffour}, we show the distribution of $(1-r)/(1+r)$ for various
polarization states of the $t \bar t$ pair.

Another method we propose is based on the fact that the moving
direction of the $l^+$ and $l^-$ are strongly correlated in the
center--of--mass frame of the $t \bar t$ pair. To illustrate this
point, we show in \Fig{ffive} the strong correlation between
$\cos\theta_{tl^+}$ and $\cos \theta_{tl^-}$, where $\theta_{tl^\pm}$ is
the angle between boost axis of the $t$ and and the momentum of the
$l^+$ ($l^-$) in the rest frame of the $t$ ($\bar t$).
The four populated corners
correspond to the four different helicity states of the $t \bar t$ pair.
Therefore, one can select individual polarization states of $t \bar t$
by making cuts on this plot. However, it is important to make sure
that the kinematic cuts won't induce an artificial CP asymmetry.
For example, one can select the $RR$ and $LL$ data sample by
making a square cut of equal area on the
$\cos\theta_{tl^+}>0$, $\cos\theta_{tl^-}>0$
and $\cos\theta_{tl^+}<0$, $\cos\theta_{tl^-}<0$ corners
without inducing an artificial CP asymmetry when studying $\A_1$.
The $\cos\theta_{tl^+}>0$, $\cos\theta_{tl^-}<0$ and
$\cos\theta_{tl^+}<0$, $\cos\theta_{tl^-}>0$ corners
are suitable to measure the form factors $A$ and $B$ from the $RL$ and
$LR$ data sample.  Similarly, information on the form factors $C$ and $D$
are contained in
the $LL$ and $RR$ samples.  The key element of this approach is to be
able to determine the moving direction of the $t$ via the decay mode of
$t \bar t \ra b l^+ \nu_l \bar b l^- {\bar \nu_l}$.  This method is therefore
useful even when some kinematic cuts are being imposed on the final observable
partons as usually done in reality for detection and triggering.
In \Fig{fsix} we show the effect of selecting various corners in the
configuration space of \Fig{ffive} on the distribution in $(1-r)/(1+r)$.
The acceptance of the kinematic cuts in selecting $LL$ or $RR$ events
is about $1\over 4$.
Different polarization states of the $t\bar{t}$ dominate in different
regions of $(1-r)/(1+r)$.  Therefore, it is possible to enhance the
probability of a particular polarization of the $t\bar t$ pair by making
cuts on $(1-r)/(1+r)$ as to be performed in the following analysis.
We also note the relative enhancement
of the $RR$ and $LL$ rates displayed in \Fig{fsix} as a result of these
selections.

\subsection{The Bottom Quark Ambiguity}

If the $b$ is not distinguished from the $\bar{b}$,
there are two solutions provided by this procedure because it becomes
possible to assign mistakenly the $\bar b$ with the $t$.
Nevertheless, this causes only minor difficulties.
To break the degeneracy, one first solves the kinematics given
one combination of jet momenta assignments
and then repeats the procedure for the
other combination, then one compares the two results.

We performed a simplified computation without ISR where it was not
necessary to apply both of the constraints provided by $m(W^+)$ and
$m(W^-)$, since then we know the boost of the system.  Furthermore,
with the center--of--mass energy known, we did not need to invoke the nonlinear
relations restricting the neutrinos to be massless.  This means that
given one $W$ boson mass, we were able to solve for the other.  The
tendency is that the combination where the $b$ quark was misidentified
as the $\bar{b}$ quark generates unreasonable solutions for the
$W$ boson mass that was left as a free parameter,
allowing a proper identification of the $b$ quark jets.  In
particular, it was chosen to drop any solutions that gave a negative
$W$ boson mass (this condition may be tightened).  If both solutions
yielded positive masses for the gauge bosons, they were subjected to a
further selection criteria regarding the neutrino masses.

One effect of using a value for the mass of the $W^-$
different than $\sqrt{(p(e^-)+p(\bar{\nu}))^2}$ in the
above procedure is that the solution produces magnitudes for the neutrino
momenta which are no longer equivalent to their own energies.
To remove some of the inefficiency caused by the
finite width, $\Gamma_W$, two solutions were determined.  The first
solution was obtained by fixing the mass of the $W^-$ boson to $m_W$
and solving the kinematics.
For the second solution we left the mass of the $W^-$ boson free and
fixed the mass of the $W^+$ boson to $m_W$.
An artifiact of this procedure is that
the worst solution of the two tends to provide neutrinos with larger
values of $p(\nu)^2$ and $p(\bar{\nu})^2$.  Since the neutrino is
massless, selecting the solution that gives the neutrino masses
closest to zero ameliorates the problems introduced by $\Gamma_W$ and
the remaining problem introduced by the inability to distinguish
the $b$ quark from the $\bar b$ antiquark.

We have used a Monte Carlo program to test the success rate of this
algorithm and found a 99\% efficiency in reconstructing the moving
direction of the $t$ by assuming a hundred percent detection
efficiency of the energy of the $b$ and $\bar b$. About 0.5\% of the
solutions misidentified $b$ quarks as $\bar b$ antiquarks and the other
0.5\% consisted of kinematic configurations which could not provide a
solution given our criteria.  To determine whether one can measure CP
violation effects better than a percent from this method requires
further study to ensure that no artificial CP violation is induced
by removing from consideration those events for which no solution was found.
Attempting to account for some detector
inefficiency, this procedure was also studied with the inclusion of
a gaussian smearing of the momentum for the $b$ and $\bar b$ with
$\Delta E(b)/E(b)=0.5/\sqrt{E(b)}$ and for the visible leptons with
$\Delta{E(l\pm)}/{E(l\pm)}=0.15/\sqrt{E(l\pm)}$.
The results in this case gave
a $96\%$ efficiency in assigning the bottom momenta to the proper top
quarks where $3.3\%$ of the time this assignment was incorrect and
$0.7\%$ of the events provided no solution given our criteria.
It is expected that the efficiency in removing the
bottom quark ambituity should improve slightly
by selecting the best
values of $p(\nu)^2$ and $p(\bar{\nu})^2$
among a range of choices where we do not restrict ourselves
to fixing the $W$ boson masses to $m_W$, but rather allow ourselves to
check a variety of solutions for $W$ boson masses within, e.g.,
$m_W\pm 2\Gamma_W$.  Note that in distinguishing the $b$ from the $\bar b$
through their kinematics as we have described, we have not taken
advantage of other information which can be made available such as
the charge of the leading particle in the bottom quark jet.\cite{strikman}

Summarizing, we described how the top quark analyzes its own
polarization through its decay $t\to bW^+\to b\l^+\nu$ such that a
moving top (anti)quark with right--handed (left--handed) helicity
tends to produce the $l^+$ ($l^-$) with its momentum aligned with the
top (anti)quark direction of motion.  We have demonstrated that the top
quark momentum can be determined well including smearing and width
effects with very little ambiguity in deciding which bottom flavored
quark goes with which top flavored quark.  If the CP violation effects
are very small, however, the error introduced by the ambiguity in
distinguishing the $b$ from the $\bar b$ may interfere with the measurement.

A $1\,$TeV machine can do better than a $500\,$GeV machine regarding
the determination of the top quark momentum, polarization, and
which bottom quark should be associated with the $t$, all
due to the larger boost the top quark receives.

\section{Conclusion}

We have shown how to determine the top quark momentum in \dilep\ and
\ljets\
by solving for the momenta of the neutrinos, including the effects
from initial state radiation.  Assuming the mass of the top quark
is known, an accurate
determination of the top quark moving direction was obtained, even
when including detector effects by smearing the particles' momenta
and taking into account the finite width effects from the $W$ boson.
The efficiency
of this algorithm was affected very little by the degeneracy created
by treating the $b$ and $\bar{b}$ as indistinguishable, and it is possible
to optimize the efficiency to account for width effects in the mass of the
$W$ boson.  With the top quark direction known, the door is open to
performing studies using the polarization of the top quark.

Through angular correlations between the lepton and top quark
directions, we have shown how to separate the different polarization
states for the $t\bar{t}$ pair.  Applying these tools, we have
examined how CP violation effects and form factor magnitudes may be
studied in \dilep\ and concluded that large luminosities
($3\times 10^4\,\hbox{fb}^{-1}$)
are required to measure
a CP violation asymmetry $\A_1$ at the order of $10^{-2}$ at the NLC.
On the other hand,
the form factors can be measured to some accuracy.
For the $l^\pm +\hbox{jets}$ mode at a $500\,$GeV machine ,
the results indicate that within the $68\%$ ($90\%$)
confidence limit, it should be
possible to find $F_1^{(L)}$ to within about $3\%$ ($8\%$),
while
$F_1^{(R)}$ can be known to within roughly $5\%$ ($18\%$).
In this case the $F_2$ form factors
can be known to within about
$0.0\pm 0.004$ ($0.0\pm 0.02$) at a confidence level of $68\%$ ($90\%$).
Isolating the $RL$
and $LR$ contributions can improve the precision of the measurements of these
form factors. For instance, we saw that in the dilepton mode
that $F_1^{(L)}$ can be measured to about $6\%$ at the $68\%$ confidence level
using unpolarized beams, yet by
selecting the $RL$ contribution using the methods discussed in this
paper, it appears possible to improve this bound to $4\%$ in this mode.

A $1\,$TeV machine can do better than a $500\,$GeV machine in
determining $F_1^{L,R}$ because the relative sizes of the $RR$ and
$LL$ production rates become small and furthermore, the top quark is
boosted more in a $1\,$TeV machine thereby allowing a better
determination of its direction.  A $1\,$TeV machine makes the CP
asymmetry measurement more difficult, however, because of the smaller
rates in $RR$ and $LL$.

\vbox{
\vskip 0.5cm
{\bf Acknowledgements}
\vskip 0.5cm
The authors wish to thank D.~Burke, Darwin Chang, K.~Fujii, Gordon Kane
and Michael Peskin for useful discussions.  This work was funded in part by
the TNRLC grant \#RGFY9240.}

\vfill\eject
%
%
\nosechead{References}

\ListReferences

\vfill\eject
\nosechead{Table Captions}
\item{Table 2.1}{Born level production rates for \eett\
at $\sqrt{s}=500\,$GeV and $1\,$TeV using $m_t=140\,$GeV listed according to
$t$ and $\bar t$ polarizations.}
\item{Table 4.1}{The upper and lower bounds indicate the statistical
range within which we can
determine the form factors from the angular distribution of the top quark
in \dilep\
given a $50\,\hbox{fb}^{-1}$ luminosity at the NLC within a $68\%$ confidence
level.}

\item{Table 4.2}{The upper and lower bounds indicate the statistical
range within which we can
determine the form factors from the angular distribution of the top quark
in \dilep\
given a $50\,\hbox{fb}^{-1}$ luminosity at the NLC within a $90\%$ confidence
level.}

\item{Table 5.1}{The efficiency in solving for the top quark momentum
using initial state radiation, smearing, and a Breit--Wigner width
on the $W$--bosons is described here by three values:
the percentage of solveable configurations; the
cosine of the angle between the true top quark direction and the direction
given by solving the kinematics ($\zeta\equiv\cos(t_{true},t_{solved})$);
the root mean squared of $\zeta$.}

\vfill\eject
\nosechead{Figure Captions}

\item{Figure 3.1}{This plot displays the variation of the asymmetry $\A_1$ in
\Eq{eseven} with the mass of the Higgs boson for Weinberg's model using
$\sqrt{s}=500,1000\,$GeV and $m_t=140\,$GeV.}

\item{Figure 3.2}{This plot displays the variation of the asymmetry $\A_1$ in
\Eq{eseven} with $\sqrt{s}$ for Weinberg's model using $m_t=140\,$GeV
and $m_H=0.1,0.4,1.0\,$TeV.}

\item{Figure 4.1}{In $d\sigma /d\cos\theta$ we see the dependence of the
production rates for polarized $t\bar{t}$ states on the top quark
polar angle.  This angle $\theta$ is measured with respect to the
incoming $e^-$ beam which is taken to move along the $+\hat{z}$
direction.}

\item{Figure 4.2}{Same as \Fig{fone} except for a purely left--handed $e^-$
beam.}

\item{Figure 5.1}{The center--of--mass energy of the $e^-e^+$ annihilation when
ISR is taken into account.  }

\item{Figure 5.2}{The boost, $d\sigma /d\beta$ vs.
$\beta={\vec{\bf p}(t)+\vec{\bf p}(\bar{t})\over E(t)+E(\bar{t})}$,
as generated by ISR.
}

\item{Figure 5.3}{The angular distribution
$d\sigma /d\cos\theta$ vs. $cos\theta$ as dictated by the top momentum
provided by the Monte Carlo compared against the top momentum found
when solving the kinemtics incuding the effects of ISR.  }

\item{Figure 6.1}{Helicity arguments control the preferred moving
direction for the $e^+,e^-$ as they are produced in the respective
rest frames of the $t,\bar{t}$ decays ($t\to be^+\nu$, $\bar{t}\to
\bar{b}e^-\bar{\nu}$).  The subscript $R$ ($L$) indicates
right--handed (left--handed) quark helicity.  }

\item{Figure 6.2}{$E(l^+)$ and $E(l^-)$ distributions for \eett\ when
both final state top quarks have identical helicities.  }

\item{Figure 6.3}{The distribution for the production of polarized $t\bar{t}$
pairs, $d\sigma /d[(1-r)/(1+r)]$ vs. $(1-r)/(1+r)$ for $r\equiv
E(l^+)/E(l^-)$.  }

\item{Figure 6.4}{These plots present the angular correlation between the
charged leptons produce from the polarized $t$ and $\bar t$ decays in
\dilep.  Boosting to the individual center of mass frame for the
decays of the $t$ and $\bar t$, we plot the cosine of the angle
between the momentum of each charge lepton and the boost axis as
defined by the top momentum.  Note how each combination of $t$,$\bar
t$ helicities (RR,LL,RL,LR) occupy separate quadrants.  }

\item{Figure 6.5}{The distribution for the production of polarized
$t\bar{t}$ pairs, $d\sigma /d[(1-r)/(1+r)]$ vs. $(1-r)/(1+r)$ for
$r\equiv E(l^+)/E(l^-)$ when angular cuts are applied to the final
state $l^+$ and $l^-$ in the respective decay frames of the $t$ and
$\bar{t}$.  }

\vfill\eject
\PrintTables
\vfill\eject
\PrintFigures

\bye